\documentclass[twocolumn]{aastex63}
\usepackage{inputenc}
\usepackage{comment}
\usepackage{colortbl}
\usepackage{graphicx}
\usepackage{soul}
\usepackage{hyperref}
\usepackage{multirow}
\usepackage{animate}

\newcommand{\chandra}{\textit{Chandra}}

\received{99 Doc 2099}
\revised{99 Doc 2099}
\accepted{99 Doc 2099}
\submitjournal{ApJ}

\shorttitle{15 Years @ M31*}
\shortauthors{DiKerby, Zhang, \& Irwin 2025}
\graphicspath{{./}{figures/}}
\begin{document}

\title{Fifteen Years of M31* X-ray Variability and Flares}

\author[0000-0003-2633-2196]{Stephen DiKerby}
\affiliation{Department of Physics and Astronomy \\
 Michigan State University,
East Lansing, MI 48820, USA}

\author[0000-0002-2967-790X]{Shuo Zhang}
\affiliation{Department of Physics and Astronomy \\
Michigan State University, East Lansing, MI 48820, USA}

\author[0000-0003-4307-8521]{Jimmy Irwin}
\affiliation{Department of Physics and Astronomy \\
University of Alabama, Box 870324
Tuscaloosa, AL 35487 USA}
\affiliation{Eureka Scientific, Inc. \\
2452 Delmer Street, Suite 100
Oakland, CA 94602, USA}

\begin{abstract}

We append an additional fifteen years (2009-2024) to the \chandra\ X-ray light curve of M31*, the supermassive black hole at the center of M31, the Andromeda galaxy. Extending and expanding on the work in \cite{Murmur1}, we show that M31* has remained in an elevated X-ray state from 2006 through at least 2016 (when regular Chandra monitoring ceased) and likely through 2024, with the most recent observations still showing an elevated X-ray flux. We identify one moderate flare in 2013 where the other nuclear X-ray sources are in low-flux states, making that flare a valuable target for followup with multiwavelength and multimessenger archival data. We extract a mostly uncontaminated spectrum for M31* from this observation, showing that its X-ray properties are similar to those observed at Sgr A* in its quiescent state by \cite{ChandraSagA}. Furthermore, we find no substantial change in the source's hardness ratio in the 2006 and 2013 flares compared to the post-2006 elevated state, suggesting the these flares are increases in the regular X-ray emission mechanisms instead of entirely new emission modes. Our extended light curve for M31* provides valuable context for multimessenger or multiwavelength observations of nearby supermassive black holes.

\end{abstract}


\section{Introduction and Context}
\label{sec:Intro}

Every large galaxy seems to have at its heart a supermassive black hole (SMBH), related in observationally concrete ways to the properties of the galaxy itself. In recent years, two SMBHs in particular have emerged not only as unseen gravitational forces in the crowded core of a galaxy, but as discrete objects that can be directly observed in their own right. These are Sagittarius A* (Sgr A*), only 8 kpc away in our own galaxy, and M87*, a thousand times more distant in the elliptical galaxy M87 but a thousand times more massive; these SMBHs have been directly imaged by the Event Horizon Telescope \cite{EHT1,EHT2} and followed up extensively with multiwavelength observations.  These direct investigations of local SMBHs have tested fundamental theories like Einstein's relativity and have opened new avenues for probing the relationship between a galaxy and its black hole.

Sgr A* and M87* represent opposite ends of SMBH behavior.  M87*, with a mass of $\approx 6 \times 10^9 M_\odot$ \citep{GebhardtM87Mass}, powers an AGN and blasts a jet of relativistic material into intergalactic space \citep{M87Jet,Thim2024ChandraProper,Sheridan2024M87Jet}. Sgr A*, on the other hand, has a mass of only $\approx 4 \times 10^6 M_\odot$ \citep{GhezSagAOrbits,SagAmass}, and when not displaying radio, IR, or X-ray flares \citep{SagAIR,SagAXray,SagAradio} it is generally in a quiescent state with luminosity far below the Eddington limit, even compared to other ``low-luminosity AGN''. In particular, the detection of an X-ray source coincident with the radio position of Sgr A* \citep{SagAXray} has made Sgr A* a substantial target for followup X-ray observations, including with higher-energy telescopes including NuSTAR.  In \cite{SagAXray}, the authors noted that the X-ray source coincident with Sgr A* has a soft spectrum modeled as a power-law with $\Gamma_X = 2.7$ or as an optically thin thermal plasma with $k_B T = 1.9 \:\rm{keV}$ during quiescent states. \cite{Zhang2017} demonstrated that Sgr A* X-ray flares have a harder spectral index of $\Gamma_X \approx 2$. In these ways Sgr A* and M87* while having very different masses and absolute luminosities, are both characterized by substantial variability in multiple bands.

There is another SMBH that can feasibly be spatially resolved from its surroundings without the use of a planet-spanning interferometer; M31* is the black hole at the center of M31, the Andromeda galaxy, the nearest large galaxy to our own. The inner nucleus of M31 has a optical/IR double nucleus \citep{M31DoubleCore}, the two sources being named P1 and P2, which largely align with the position angle of M31's disk as a whole. This double nucleus is understood as populations of stars and gas near the SMBH \citep{M31NucleusDisk}. M31* has a mass of $\sim 10^8 M_\odot$ \citep{M31starBender}. After detections at P2 in the radio \citep{P2radio} and X-ray \citep{GarciaM31StarDetection}, P2 has emerged as the most likely location for the central SMBH. M31* has been extensively monitored in the radio with instruments like the VLA \citep{YangVLA}, providing multiwavelength context for energetic processes in the nuclear region of M31.

\begin{figure}
    \centering
    \includegraphics[width=\linewidth]{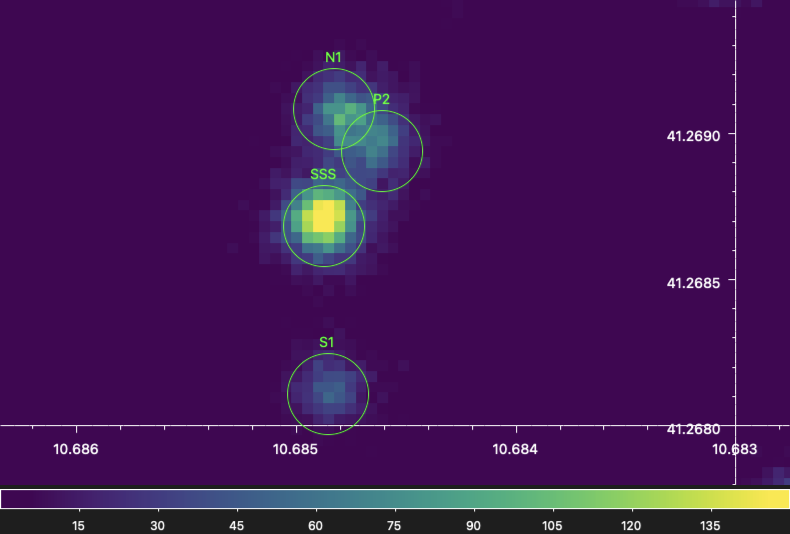}
    \caption{The summed, aligned \chandra\-HRC observations showing the four nuclear sources close to M31*. P2 is the source associated with M31*, while N1, SSS, and S1 are nearby nuclear sources contributing substantial PSF contamination. The colorbar is linear with respect to total events.}
    \label{fig:all_field}
\end{figure}

X-ray analysis of M31* is hampered by the crowded field of X-ray sources in the nucleus of M31; in particular, there are four X-ray sources nestled within a few arcseconds of the SMBH. Using the terminology from \cite{P2XrayFirst} and \cite{LiOverallM31}, these are P2 (spatially coincident with the radio SMBH position), N1 only $\sim 0.5"$ east of P2, SSS about $\sim 1"$ south, and S1 another $\sim 1"$ further south. The relative positions of these four sources are shown in Figure \ref{fig:all_field}, a \chandra\-HRC summed count map produced following the processing described below.

The only currently operative X-ray telescope with the spatial resolution necessary for resolving these sources is the \chandra\ X-ray Observatory \citep{ChandraOverview}, either with sub-pixel event reconstruction with the Advanced CCD Imaging Spectrometer (ACIS) or with the fine precision afforded by the High Resolution Camera (HRC). Other observatories can image the entire nuclear region, but the busy field near the SMBH means that it is impossible to disentangle the emission of each individual source without high spatial resolution. Using 2D spatial analysis, \cite{Murmur1} detached the X-ray emission of P2 from its neighbors and produced a 10-year light curve for P2 from 2000 to 2010. It is worth noting that all four nuclear X-ray sources close to the SMBH show substantial X-ray variability in \cite{Murmur1}. To create a light curve of P2 requires dealing with four sources with overlapping point spread functions (PSFs) and confounding variability.

In \cite{Murmur1}, the authors noted that before 2006, M31* was in a quiescent state, with an average ACIS-I $0.5 - 8 \:\rm{keV}$ count rate of $(0.3 \pm 0.1) \times 10^{-3} \:\rm{cts/s}$. In 2006, a single ACIS observation recorded a substantially elevated flux of $(77 \pm 6) \times 10^{-3} \:\rm{cts/s}$, interpreted as an X-ray flare. Subsequent observations, starting several weeks after the flare and continuing to 2010, showed an elevated average flux of $(6.9 \pm 0.3) \times 10^{-3} \:\rm{cts/s}$ with greater variability as well, well above the low state observed before the 2006 flare.

\begin{figure*}
    \centering
    \includegraphics[width=\textwidth]{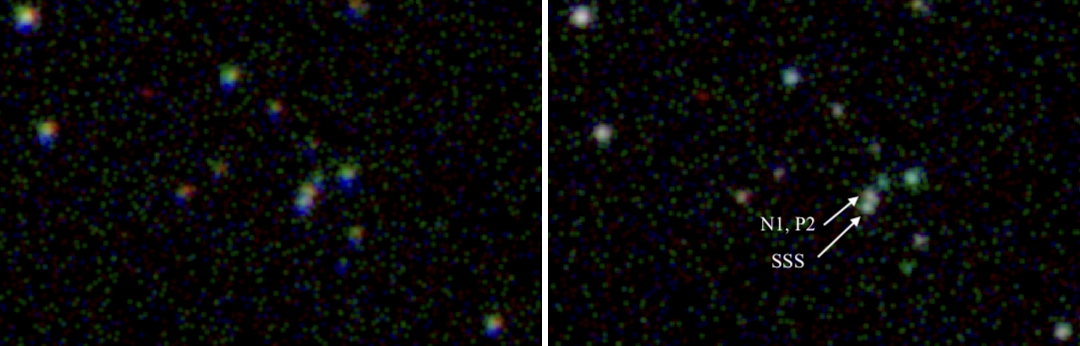}
    \caption{\chandra\ X-ray events near the nucleus of M31 in HRC-mode observations 13178 (red), 27695 (green), and 28911 (blue) before re-aligning of data (left) and after (right). Before realignment, 27695 has an eastward bias compared to the other observations, and 28911 has a southerly bias. After correction, the biases are removed and the centroids of each color are the same, though inter-observation variability is still visible as net coloration of each source.}
    \label{fig:align_step}
\end{figure*}

It has been thirteen years since \cite{Murmur1}, and \chandra\ is still the only X-ray telescope with the resolving power to dissect the crowded X-ray environment around the nucleus of M31. With the advent of direct imaging of SMBHs \citep{EHT1,EHT2},  the discovery that M31* can undergo X-ray flares that have long-lasting effects on its behavior, and the emergence of flares from SMBHs as exciting candidates for particle astrophysics experiments like IceCube, it is worth revisiting the past $\sim 15$ years of \chandra\ observations aimed at the nucleus of M31.

In this work, we extend and expand the analysis from \cite{Murmur1} to include fifteen years of \chandra\ observations, from 2009 to 2024.  In Section \ref{sec:ObsProc} we describe the observations obtained for this work and the pipeline used to prepare data for analysis. In Section \ref{sec:ana} we describe our analysis of the generated light curve and spectral analysis of a fortuitous observation. Finally, in Section \ref{sec:Conc} we present our conclusions and next steps.

\section{Observations and Processing}
\label{sec:ObsProc}

We downloaded all \chandra\ observations in the public Chandra Data Archive targeting the nucleus of M31 from 2009-01-01 to the present, plus several recent observations that are still proprietary, totaling $638 \:\rm{ks}$ of HRC data across 37 observations and $576 \:\rm{ks}$ of ACIS data across 68 observations. We conducted all processing and analysis using \verb|ciao| and \verb|sherpa| version 4.16 \citep{CiaoDoc}. 

We used the standard Chandra reprocessing to reprocess the primary and secondary products. For ACIS observations, we used \texttt{fluximage} to obtain flux, exposure, and PSF maps for each ACIS observation in the \texttt{broad} band with spatial rebinning factor of $0.2$ for a spatial binsize of $0.2 \times 0.4920 \arcsec = 0.098 \arcsec $.  For HRC observations, we used \texttt{fluximage} to obtain the same maps in the \texttt{wide} band ($\approx 0.1 - 10.0 \:\rm{keV}$, using events with $0 < \rm{PI} < 254$ as in the Chandra Source Catalog \citep{ChandraSourceCatalog1}) with no spatial rebinning for a spatial binsize of $0.13 \arcsec$.

\begin{deluxetable}{lll}
\tablecaption{The J2000 (fk5) positions of the four primary sources modeled in the nucleus of M31*, following the definitions used in \cite{LiOverallM31} and \cite{P2XrayFirst}.} \label{tab:posPrim}
\tablewidth{\textwidth}
\tablehead{
\colhead{Source name} & \colhead{RA} & \colhead{Dec} }
\startdata
S1 & 00 42 44.36 & +41 16 05.2 \\
SSS & 00 42 44.37 & +41 16 07.3 \\
N1 & 00 42 44.36 & +41 16 08.7 \\
P2 & 00 42 44.31 & +41 16 08.2\\
\enddata
\end{deluxetable}

It is necessary to recalibrate the spatial coordinates of each observation to ensure that the positions of the four nuclear X-ray sources are stable between observations. We use \verb|celldetect| to locate the X-ray sources present in each observation, then \verb|wcs_match| and \verb|wcs_update| to compare these detections with the position of X-ray sources in the \chandra\ Source Catalog - 2nd Edition \citep{CSC2}. In this way we subtly shift the world coordinate systems of each observation to produce a corrected events file. This step guarantees that previously used positions for P2 can be used for all modeling and fitting steps without further complications, as shown in Figure \ref{fig:align_step}. After realigning each observation, we generate a new events file and run \verb|fluximage| again to produce aligned flux, exposure, and PSF maps.

% This line would allow you to embed a nice movie GIF of the images, but would require Adobe or something to show that in a PDF.
%\animategraphics[loop,controls,width=\columnwidth]{2}{./figures/animation/anim_}{0}{32}

Using the definitions in \cite{LiOverallM31} and \cite{P2XrayFirst}, Table \ref{tab:posPrim} gives the coordinates of the four nuclear sources used in subsequent analysis. The re-aligned $0.5-8 \:\rm{keV}$ count maps produced by \verb|fluximage| are the primary files used to model the fluxes of each individual source in each observation.

\section{Analysis}
\label{sec:ana}

\subsection{Light Curve Construction}

\begin{figure*}[t]
    \centering
    \includegraphics[width=0.9\linewidth]{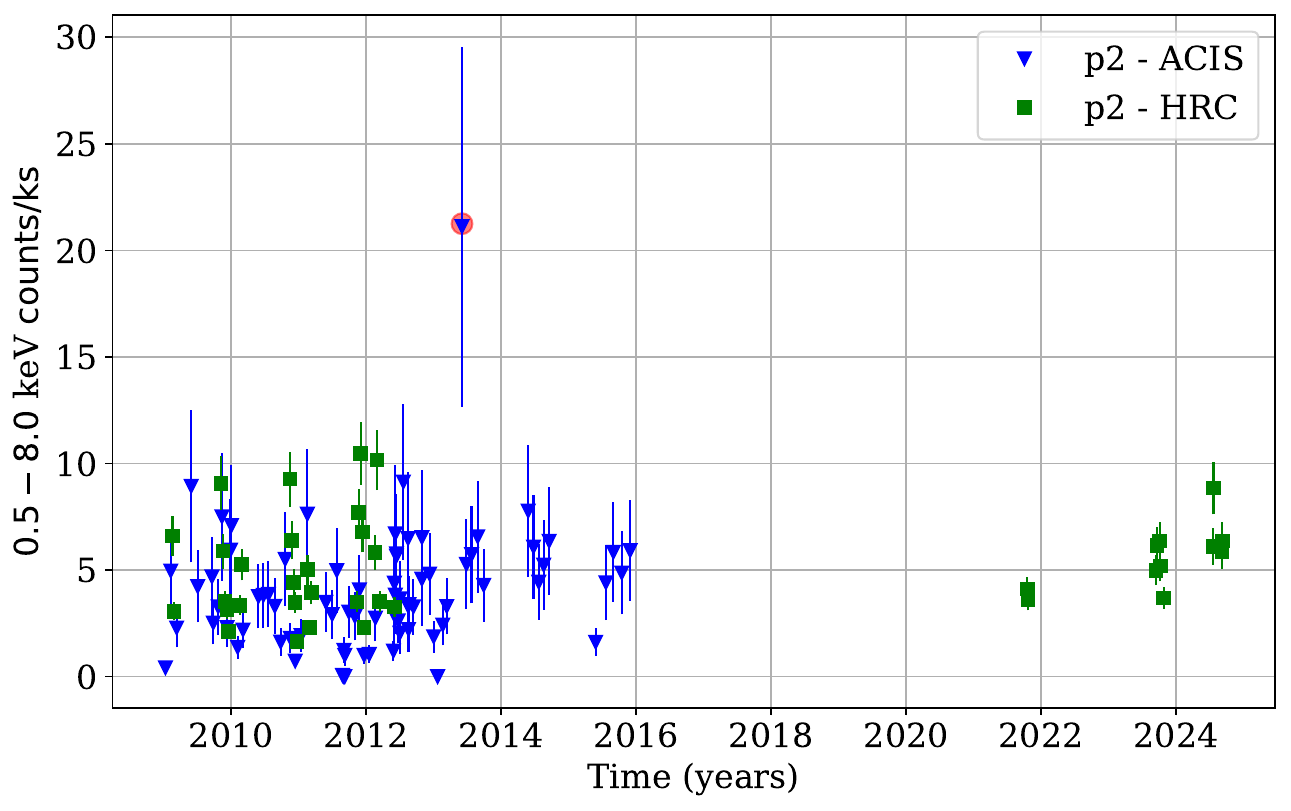}
    \caption{The ACIS-I equivalent count rate light curve for P2 from the past fifteen years, including ACIS (blue arrows) and HRC (green boxes) observations. Observation $15324$ in 2013 with an X-ray flare is highlighted in red.}
    \label{fig:P2Curve}
\end{figure*}

Because of the overlapping PSFs of the four nuclear sources in Table \ref{tab:posPrim} and the independent variability of each source, we obtain the count rate from P2 by simultaneously modeling the 2D flux distribution from the four nuclear sources in \verb|sherpa|. Loading in the realigned count map of each observation, we focus on an $11"$-diameter circle containing only the four nuclear sources. Because this sky region is small compared to the entire ACIS or HRC fields, we assumed that the PSF and exposure maps are effectively constant across the region, fixed at the value of the center of the field, which holds to sub-$1\%$ precision for all our observations.

Our 2D model for the sky region is a constant background plus four delta-function point sources at the positions of the four nuclear sources, each delta function being convolved with a \verb|beta2d| model to parameterize the PSF of the telescope. We use a numerical model instead of the empirical PSF used in \cite{Murmur1} to avoid biases introduced by the variable PSF across the \chandra\ field of view, to avoid confounding effects from the densely packed other X-ray sources used to create the empirical PSF in \cite{Murmur1}, and to allow for direct integration of total counts. We use a \texttt{beta2d} model instead of a Gaussian after testing a Gaussian PSF and finding systematic biases in the wings of the PSF in modeling the nuclear sources and in the total counts upon integrating the Gaussian PSF.

Using other non-nuclear sources located within a few arcseconds of P2, we find that the PSFs of the nuclear sources are well fit with $\alpha = 2.5$ (the polar integral of \texttt{beta2d} converges for $\alpha > 1$) and $r_0$ between 4 and 5 pixels (depending on the \chandra\ observation given the local PSF at the nuclear sources), the only free parameters in our 2D modeling are $N_0$ an amplitude linearly related with the total counts of each independent source.

We run the fit and obtain confidence limits using the \verb|fit| and \verb|confidence| commands, using Nelder-Mead optimization \citep{NelderMead} and the Cash statistic as optimization metric due to low photon counts in individual bins \citep{CashStat}. We calculate the total number of photons for each source using the polar form of \verb|beta2D|, given by

$$ \rm{counts} = N_0 \int_{r=0}^{\infty} \left( 1 + \left( \frac{r}{r_0} \right)^2 \right)^{-\alpha}  2 \pi r \: dr $$

\noindent for each of the nuclear sources in each observation. Then, we divide by the detector livetime in each observation to obtain a modeled count rate. 

Because of the differences in effective area between ACIS-I, ACIS-S, and HRC-I observations, we apply a scaling factor to convert ACIS-S and HRC-I count rates into equivalent ACIS-I $0.5-8.0 \:\rm{keV}$ count rates. For this rescaling, we assume an incident power law spectrum with $\Gamma_X = 1.7$ and $n_H = 10^{21} \:\rm{cm^{-2}}$ as in \cite{Murmur1} and use the range and central energies definitions for the ACIS \texttt{broad} and HRC \texttt{wide} bands from \cite{ChandraSourceCatalog1}. By conducting this conversion, we match the units and methods used for Figure 3 from \cite{Murmur1}. We show the light curve produced for P2 in Figure \ref{fig:P2Curve}.

\begin{figure}
    \centering
    \includegraphics[width=\linewidth]{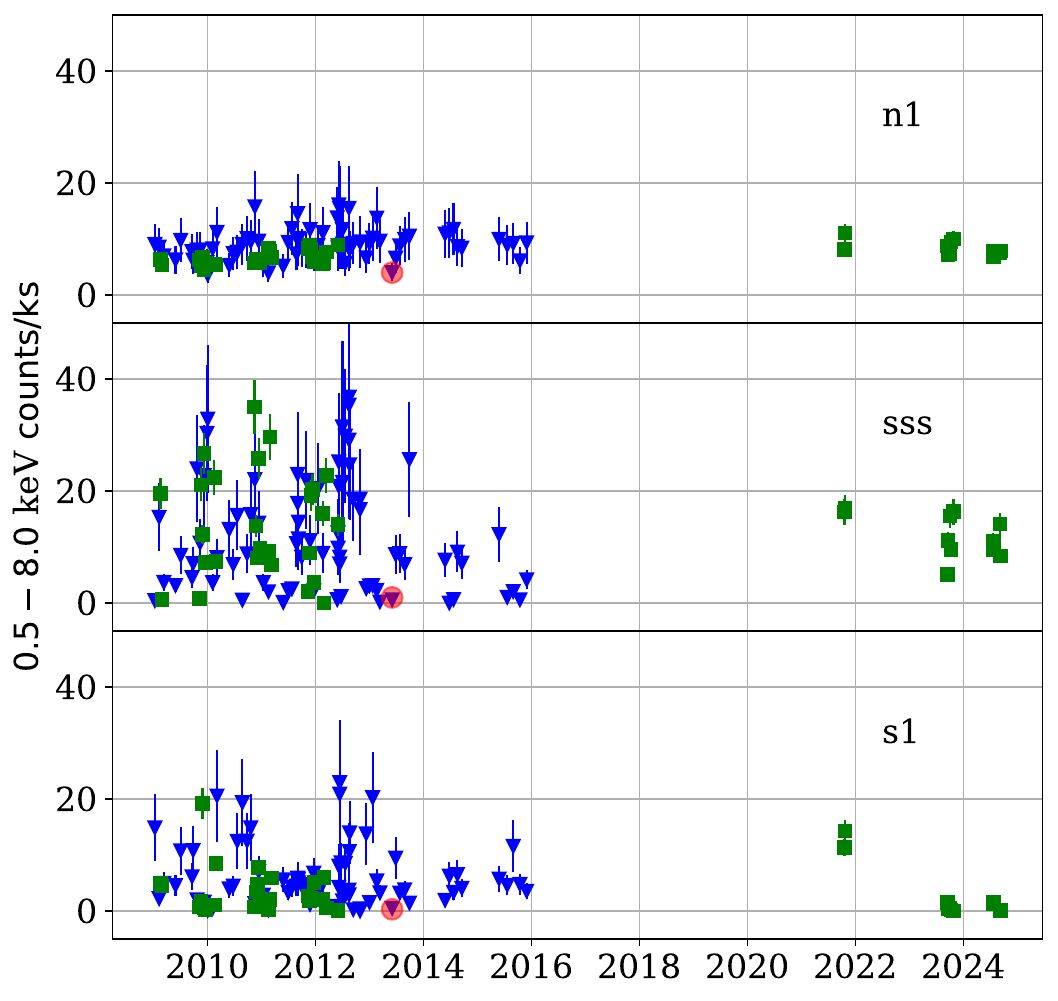}
    \caption{Light curves for the other three nuclear X-ray sources close to M31* on the same scales as Figure \ref{fig:P2Curve}, for N1 (top), SSS (middle) and S1 (bottom). Observation $15324$ in 2013 with an X-ray flare is highlighted in red.}
    \label{fig:OtherCurve}
\end{figure}

Exposures with ACIS that are too brief to establish suitable fits for the four nuclear sources (with duration $\sim 200 \:\rm{s}$ or less, taken simultaneously to account for pile-up of transients) result in less than a single photon from P2 per observation and do not constrain the count rates from the other nuclear sources. We exclude them from analysis.

In producing this light curve, we also obtain flux histories for the other three nuclear sources. While not immediately applicable to this work, examining these trends is relevant, especially to ensure that the fluxes modeled for N1 and P2 display no correlation that would imply degeneracy between their interlocked PSFs. Figure \ref{fig:OtherCurve} shows the light curves of N1, SSS, and S1 over the same timescales as Figure \ref{fig:P2Curve} for P2. Conspicuously, SSS and S1 show a substantially higher flux and inter-observation variability ($\sigma_F / \bar F \approx 1$) than N1 or P2 ($\sigma_F / \bar F \approx 0.5$). We find no correlation between the fluxes of any of the four nuclear sources between observations, suggesting that our analysis procedure is successfully and independently fitting the fluxes of the four nuclear sources.

\subsection{Spectral Analysis of P2}

Extracting a clean spectrum for P2 is complicated by the overlapping PSFs of the other nuclear X-ray sources. However, one important feature of the four nuclear sources could facilitate mostly uncontaminated fitting; the variability of the nuclear sources means that if the other nuclear sources are all in low-flux states, the X-ray emission at the position of P2 could be almost entirely from P2 itself.

\begin{figure}
    \centering
    \includegraphics[width=\linewidth]{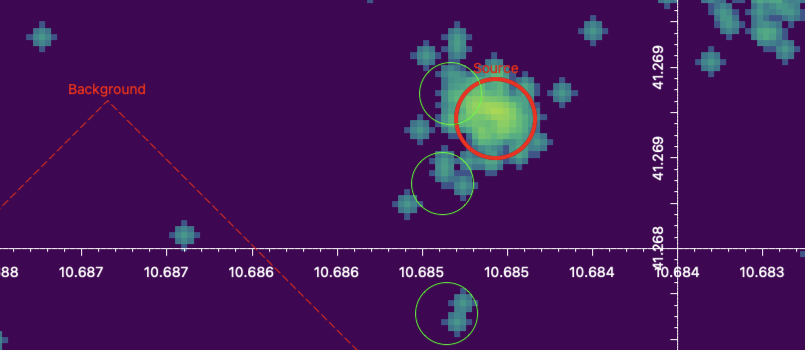}
    \caption{Observation 15324 captures every nuclear source except P2 in a low state (green circles), allowing for uncontaminated spectral fitting. The source region (red solid) and background region (partial red rectangle) for spectral analysis are also shown.}
    \label{fig:15324}
\end{figure}

To catch P2 in a hopefully uncontaminated state, we manually examine the flux maps for each observation and the lightcurves for each source in Figure \ref{fig:P2Curve} and \ref{fig:OtherCurve}. We discover Obs. $15324$, where S1, SSS, and N1 have count rates of $0.49$, $0.59$, and $4.5$ $ \:\rm{cts/ks}$, respectively, each substantially below their average fluxes, highlighted in red in Figure \ref{fig:OtherCurve}. In a doubly lucky coincidence, P2 has a notably elevated count rate of $21.1 \:\rm{cts/ks}$, four times its average post-2006 rate, making this event a possible second flare from M31*. Given the above count rates, P2 is contributing over 80\% of the photons from P2+N1. We approach spectral modeling of this region with caution, as $\sim 20 \%$ of the photons may originate from the nearby N1.

We create a circular source region centered at the position of P2, shown in red in Figure \ref{fig:15324}, and a large rectangular background region shown partially in Figure \ref{fig:15324} that contains no other bright sources. Examining the photon arrival times in the source region, we see no evidence for count rate variability at P2 in this observation, constraining the duration of this X-ray flare at M31* to above $\sim 4 \:\rm{ks}$.

We use \verb|specextract| to produce source and background spectra from the reprocessed events files for this observation, grouping the resulting source spectrum so that each bin has at least $S/N > 3$. Leaving the spectrum unbinned and using the Cash statistic \citep{CashStat} for the following fitting steps does not result in statistically significant changes in fitted parameters.

Using \verb|xspec|, we evaluate both a power-law model and optically thin thermal plasma model for the emission from P2 in this observation. The power-law model has the form \verb|tbabs * cflux * powerlaw|, with fitted parameters $n_H$ the hydrogen column density, $F_X$ the X-ray flux, and $\Gamma_X$ the X-ray slope. The thermal model has the form \verb|tbabs * cflux * raymond|. \verb|raymond| uses the model for optically thin X-ray plasma from \cite{RaymondPlasma}, for which we fix $z = 0$ and relative abundance to unity, leaving only $k_B T$ the temperature of the plasma as a free parameter. For \verb|tbabs|, we use cross-section data from \cite{Verner1996} and abundance calibration from \cite{Wilms2000}.

\begin{figure}
    \centering
    \includegraphics[width=\columnwidth]{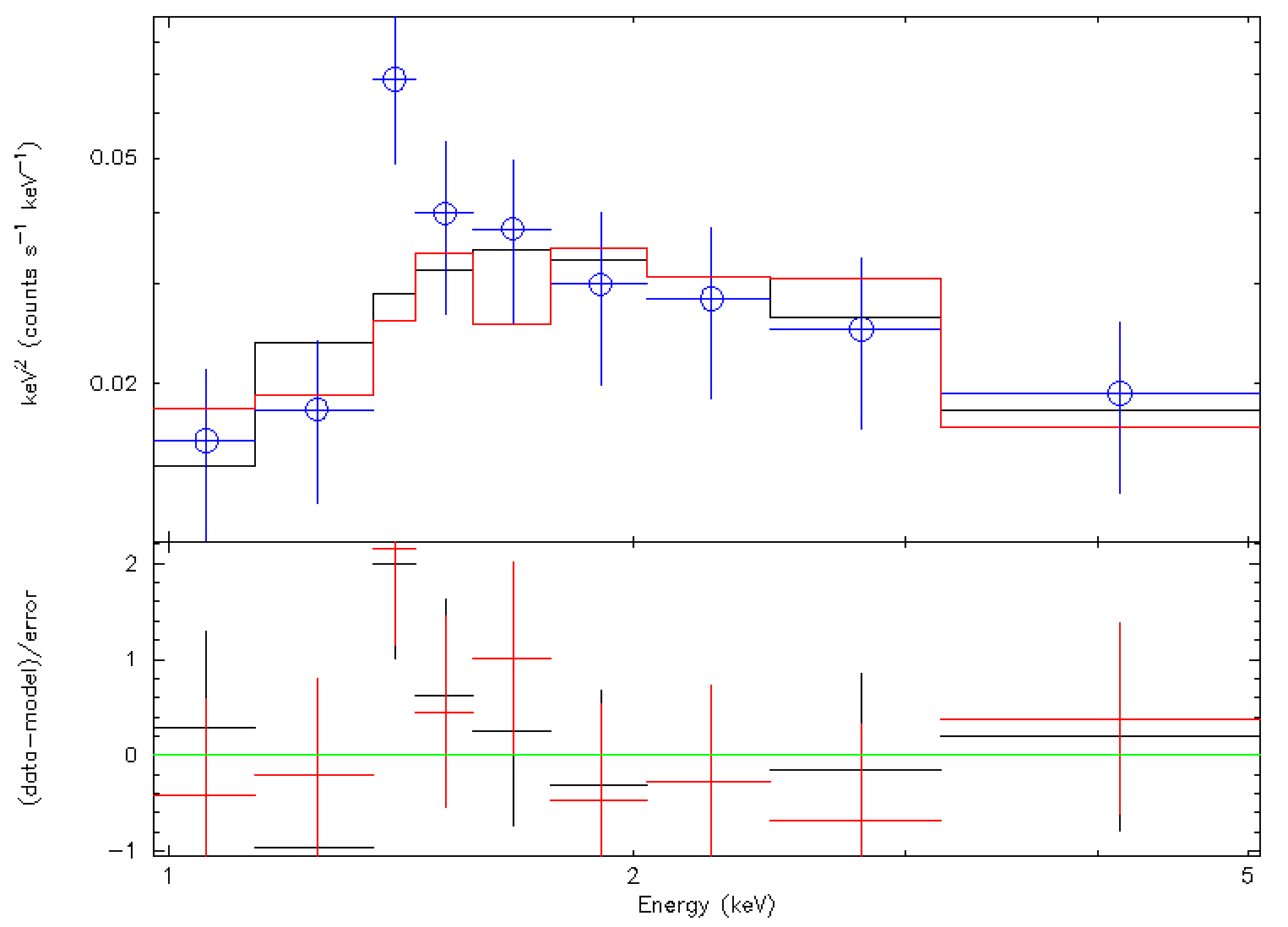}
    \caption{The grouped spectrum of P2 from observation 15324 (blue error bars and circles) with absorbed power law model (black) and thermal plasma model (red).}
    \label{fig:P2spectrum}
\end{figure}

For the power-law fit to the spectrum, we allow $n_H$ to vary upwards freely from the cataloged galactic value; $n_H > 0.17 \times 10^{22} \:/\rm{cm^2}$ from \cite{nHMap}. We find $\Gamma_X = 2.9 \pm 1.1$, $n_H = (1.1  \pm 0.6) \times 10^{22} \:\rm{cm^{-2}}$, with $\chi^2_{\rm{red}} = 0.62$. The spectrum, along with the fitted power law model with free $n_H$, is shown in black in Figure \ref{fig:P2spectrum}.

For the thermal plasma model, we once again allow $n_H$ to vary upwards from the local value to reflect the presence of gas in M31 along the line of sight. Conducting the fit using \verb|xspec|, we find $k_B T = 1.6 \pm 0.7 \:\rm{keV}$ and $n_H = (1.1 \pm 0.5) \times 10^{22} \:\rm{cm^{-2}}$, with $\chi^2_{\rm{red}} = 0.48$, suggesting another overfit. This model is shown in red in Figure \ref{fig:P2spectrum}.

\begin{figure}
    \centering
    \includegraphics[width=\linewidth]{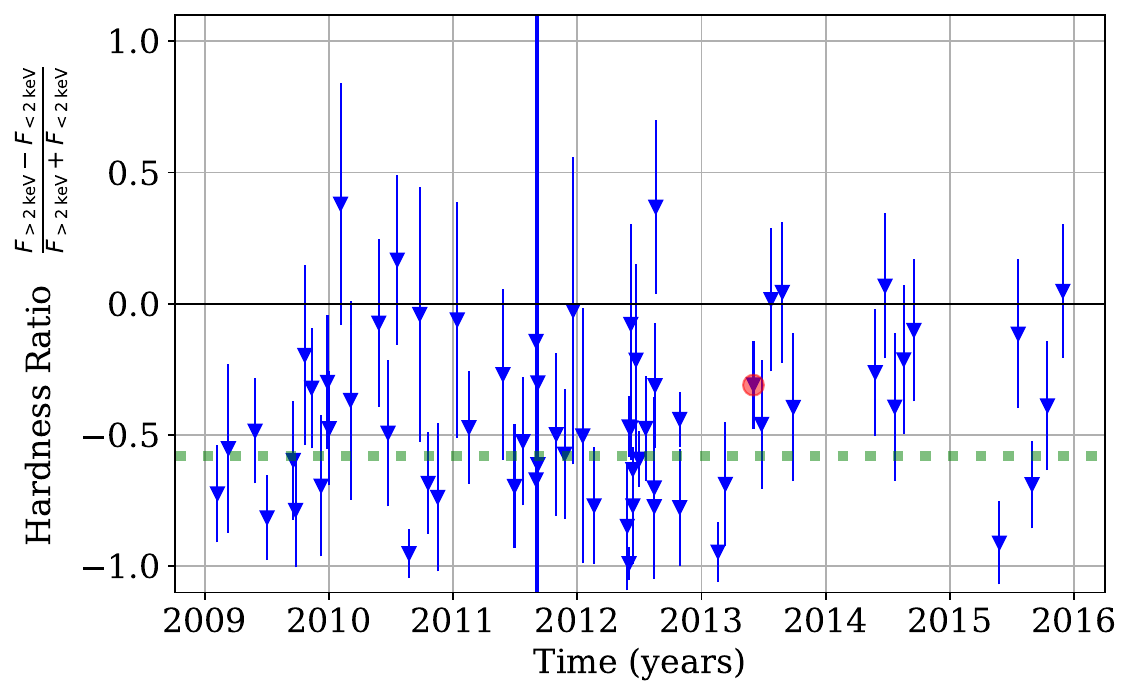}
    \caption{Hardness ratio of P2 for all the ACIS observations from 2009 to 2016 (blue triangles), showing no substantial changes in that time. The green dashed line is $HR$ for the 2006 flare (observation ID $7136$), and the red highlight is the observation in 2013 with the fortuitous flare shown in Figure \ref{fig:P2spectrum}.}
    \label{fig:HR}
\end{figure}

As an alternative approach to evaluate the spectrum of P2, we also use \verb|fluximage| to create $0.5-2.0 \:\rm{keV}$ (soft) and $2.0-8.0 \:\rm{keV}$ (hard) flux maps for each ACIS observation in order to evaluate the evolution of the hardness ratio

$$HR = \frac{ N_{\rm{hard}} - N_{\rm{soft}} }{ N_{\rm{soft}} + N_{\rm{hard}} } $$

\noindent with $N$ the predicted count rate in each band \footnote{In \cite{Murmur1}, we suspect the equation defining $HR$ has a sign error, but the reported values for $HR$ from \cite{Murmur1} seem compatible with the definition used in this work.}. Using the same procedure as before, we model the count rate in each energy range, and compute the hardness ratio $HR$ of P2 for each ACIS observation. The trend in $HR$ between 2009 and 2016 is shown in Figure \ref{fig:HR}. With low photon counts, this is a rough approach to estimating the evolving spectrum of P2 over the seven years of ACIS observation analyzed in this work.

Figure \ref{fig:HR} shows that the hardness ratio has varied around an average $HR = - 0.42$ with a standard deviation among the usable ACIS observation of $\sigma_{HR} = 0.31$. A few observations increase above $HR>0$, suggesting that for some observations P2 was producing more high-energy photons than low-energy photons, but the large uncertainties on these modeled fluxes make it difficult to evaluate the significance of these isolated incidents.

The substantial flux elevation observed in a single 2006 observation by \cite{Murmur1} (observation ID $7136$) did not occur in a fortuitous period when the other nuclear sources were ``turned off'', so it is impossible to extract an uncontaminated spectrum for P2 in that observation. However, we can compute $HR$ for that observation using the exact same analysis procedures detailed above. For that observation, we obtain $HR = - 0.58 \pm 0.22$ (shown with the horizontal dashed green line in Figure \ref{fig:HR}). For the 2013 observation examined above, we find $HR = - 0.32 \pm 0.16$, slightly higher but not outside the usual range for M31*. These values suggest that the flares in 2006 and 2013 do not have dramatically different $HR$ than when M31* is in a more quiescent period.

\section{Discussion and Conclusions}
\label{sec:Conc}

In Figure \ref{fig:P2Curve}, the X-ray light curve of P2 since 2009 shows that the equivalent ACIS-I count rate has remained elevated above pre-2006 levels at $(4.4 \pm 2.8) \times 10^{-3} \:\rm{cts/s}$ through 2016. Using energy-averaged exposure maps, this roughly converts to an X-ray luminosity of $10^{36} \:\rm{erg/s}$, assuming a distance of $750 \:\rm{kpc}$. This is roughly the same level reported in \cite{Murmur1} after the 2006 flare, suggesting that M31 has mostly remained in an elevated state since 2006 through at least 2016. Figure \ref{fig:P2Curve} shows no evidence that the elevated state existing since 2006 has ceased through the present day.

The continuous elevated state may be the consequence of the 2006 flare, with emission processes continuing unabated since then, or it may be related to numerous flares since 2006 that have gone sadly unobserved due to sparse sampling. Without additional and more frequent observations of M31's nucleus, it is not possible to qualify how frequently these flares occur, or how long their aftereffects continue. Our modeling of the hardness ratio $HR$ for P2 showed that it was largely constant between 2009 and 2016, when ACIS observations stopped, lending further credence to the continuation of the elevated state since the 2006 flare.

Notably, we do find one ACIS observation ($15324$) in 2013 with a flux of $21.1 \times 10^{-3} \:\rm{cts/s}$, brighter than during any other single observation since 2009. Examining the count rate of P2 during this observation, we find no evidence for variability on scales shorter than $\sim 4 \:\rm{ks}$. This lower limit on variability timescale does not conflict with the light-crossing time of $10^8 M_\odot$ SMBH, with 

$$\tau_{\rm{cross}} \approx \frac{3 \:\rm{km}}{c} \frac{M_{\rm{SMBH}}}{M_\odot} \approx 10^{3} \:\rm{s}$$ 

\noindent giving the lower bound on expected variability from a SMBH of mass $M_{\rm{SMBH}}$.

With the fortuitous flux reduction of N1, SSS, and S1 in Obs. $15324$, we were able to construct a rough $\sim 50$ count spectrum of P2 during a moderate flaring state (though up to $\sim 20\%$ of the photons may be contaminant from N1). Compared to the X-ray flares from Sgr A* examined in \cite{SagAXrayFlare}, the spectrum in observation $15324$ is softer -- having higher $\Gamma_X$ than the X-ray flares from Sgr A*. Furthermore, the 2013 flare only shows a $\sim 5\times$ boost in X-ray flux compared to $>10\times$ boost observed for Sgr A* flares.

Using the same power-law and thermal plasma models applied to the X-ray emission from Sgr A* in its quiescent state \citep{ChandraSagA}, we showed that both the power-law and thermal plasma models have similar $\Gamma_X$ or $k_B T$ to Sgr A* in a quiescent state \citep{ChandraSagA}, a puzzling result that suggests a contrast between X-ray activity in Sgr A* and M31*. 

The hardness ratios for the 2006 and 2013 events are not dramatically different than those for the other observations when M31* is not in a flaring state.  The fact that the mild flares from M31* have a softer spectrum than flares from Sgr A* motivates further investigation on the natures of flares from M31* and Sgr A*. In particular, the transition from cool, low-flux thermal states at M31* and Sgr A* to high-flux non-thermal states in systems like Sgr A* deserves further attention.

These pieces of evidence together hint that the 2006 and 2013 flaring events at M31* are not nonthermal X-ray flares accompanied by spectral hardening like those seen at Sgr A* \citep{SagAXrayFlare}, but are amplifications of the typical X-ray emission from the SMBH.

After the moderate flare in 2013, Figure \ref{fig:P2Curve} shows that M31* returned to the elevated post-2006 status quo through at least 2016. Further monitoring is needed to determine whether the 2006 and 2013 events are flares above a steady but elevated post-2006 emission state, or are alternatively the upper limits of flux variability for M31*. Observation $15324$ and surrounding times may also be interesting targets for multiwavelength and multimessenger archival investigations, especially since many more multimessenger observatories such as IceCube were online in 2013 compared to 2006. 

While frequent observations of M31's nucleus largely ceased in 2016, in the past few years several HRC observations have shown that P2 is still emitting at an equivalent ACIS-I count rate of $\sim 5 \times 10^{-3} \:\rm{cts/s}$. Lacking any data between 2016 through 2021, we cannot be sure that the post-2006 high state has continued through to the present day, or if M31* instead settled back into a pre-2006 low state before being re-excited by a new, unobserved flare. Given a $\sim 1\%$ duty cycle for these X-ray flares, a program of frequent \chandra\ -ACIS and -HRC observations or a monitoring campaign with a future high-resolution X-ray telescope like AXIS \citep{AXISmission} is necessary to solve the puzzle of M31*'s X-ray activity.

\software{FTools \citep{FTools}, Xspec \citep{Xspec}, Sherpa \citep{Sherpa}, CIAO \citep{Ciao}, DS9 \citep{DS9} Astropy \citep{astropy:2022,astropy:2018,astropy:2013}}

\acknowledgments

This paper employs a list of Chandra datasets, obtained by the Chandra X-ray Observatory, contained in~\dataset[DOI: 10.25574/cdc.339]{https://doi.org/10.25574/cdc.339}.

We are grateful for the various PIs who proposed observations of the nucleus of M31. While we used the resultant data for different purposes than them, we appreciate their proposals that supplied such abundant data.

\bibliography{main}{}

\end{document}